\documentclass[12pt]{article}
\begin{document}

\begin{center}
{\large \bf Interaction of super intense laser pulses with thin foil: Dopler
transformation of coherent light \\
into $X$-ray and gamma-ray bands}\\

\medskip

Vladimir A. Cherepenin (1) and Victor V. Kulagin (2) \\
\medskip
{\it (1) Institute of Radioengineering and Electronics RAS, \\
Mohovaya 18, Moscow, Russia, cher@cplire.ru \\
(2) Sternberg Astronomical Institute, Moscow State University, \\
Universitetsky prospect 13, Moscow, 119899, Russia, kul@sai.msu.ru}
\end{center}

\begin{abstract}

The formation of relativistic electron mirror produced via
ionization of thin solid target by ultraintense femtosecond laser
pulse is considered with the help of computer simulations. It is
shown that the reflection of weak counter-propagating wave from such
a mirror can produce the coherent radiation in $x$-ray and gamma-ray
bands. The spectrum of up-conversed radiation is investigated.
\end{abstract}

\par
%\medskip
\section{Introduction}

Production of coherent short wavelength electromagnetic
radiation in $X$-ray and gamma-ray bands attracts great attention
during last decades [1,2]. Different physical mechanisms have
been considered as candidates for practical implementation of
this process: generation in free-electron lasers [3-5], $X$-ray
and $\gamma $-ray lasing [1,2,6-8], high optical harmonics generation in
gases [9,10] and solids [11-13], and others. Most of the schemes
imply the use of the powerful lasers as part of the system so
the investigations received large impetus last years due to
experimental realization of ultraintense femtosecond laser
pulses [14]. In this paper we consider the Dopler up-conversion
of laser light into the $X$-ray or $\gamma $-ray bands with the help of
relativistic electron mirror produced during interaction of high
intensity ultrashort optical pulse with thin solid target in
vacuum.
\par
The idea for generation of high frequency coherent
electromagnetic radiation by Dopler transformation of the incident
wave was proposed by Landecker [15]. Actually when the light
reflects from relativistic mirror its frequency and amplitude
increase by the factor $(1+v/c)/(1-v/c)$, where $v$ is the velocity of
the mirror and $c$ is the light velocity. So for $v/c \approx  1$ the frequency
increase can be very large. Additional benefit of such scheme is its
tunability because the frequency of resulting radiation depends on
the velocity of the mirror which can be simply adjusted. The only
problem is how to produce such a relativistic mirror. Obviously it
have to be pure electronic because ordinary neutral matter cannot be
accelerated to ultrarelativistic velocities in laboratory. The
breakthrough in this problem became possible due to experimental
realization of femtosecond laser pulses of very high intensity.
During interaction of this pulse with thin foil the electrons of the
latter can be accelerated to ultrarelativistic velocities keeping
the initial geometrical form of the foil and constituting the required
relativistic mirror.
\par
Acceleration of electron bunch produced by ionization of thin
solid target with femtosecond laser pulse was considered in
details in [16]. Here in section 1 we reproduce only the results
which are important for the problem of light up-conversion, in
section 2 the process of light reflection from relativistic
electron mirror will be considered.
\par
%\medskip
\section{Formation of relativistic electron mirror}
\par
%\medskip
Let the medium is uniform in directions perpendicular to $Oz$
axis. Then it can be modeled by a set of parallel planes
(electron sheets) with constant surface density of electrons.
Each plane is supposed to have infinite dimensions in $x$ and $y$
directions. If the movement of each plane is without rotations
and deformations then all variables depend only on coordinate $z$
and time $t$ and the 1D3V model can be used for the system: the
movement of the planes can be described by three components of
velocity $\beta _{x}=V_{x}/c, \beta _{y}=V_{y}/c, \beta _{z}=V_{z}/c$
and one coordinate $Z$ [17]. In
present paper the analytical - numerical variant of 1D3V model
is used which reduces to the system of ordinary differential
equations with delay.
\par
Charge density and current are described by the following
formulas for the electron sheet ($\sigma $ is a surface charge density)
\par
\begin{equation}
\label{eq1}
\rho (z,t) = \sigma
\delta (z - Z(t)) \qquad {\mathbf{j}}(z,t) = \sigma
{\mathbf{v}}(t)\delta (z-Z(t))
\end{equation}

\noindent where $Z(t)$ is $z$ coordinate of a sheet. Then the
solutions of Maxwell
equations for the radiation fields of the medium at coordinate $z$
and time $t$ can be obtained with the help of Green function and
have the form [18]
\par
\begin{eqnarray}
\label{eq2}
 E_{z}(z,t) & = & 2\pi \sigma
{\mathrm sign} (z - Z(t')) \nonumber
\\
 {\mathbf{E}}_{ \bot } (z,t) & = & -2\pi \sigma
{\beta _{ \bot}(t')\over {1 - \beta _{z}(t'){\mathrm
sign}(z - Z(t'))}}
 \\
 {\mathbf{H}}(z,t) & = & 2\pi \sigma
{\mathrm sign} (z - Z(t')) {{\left[ \beta _{ \bot}
(t'),{\mathbf{e}}_{z} \right]} \over {1 -\beta _{z} (t') {\mathrm
sign} ( z - Z(t'))}} \nonumber
\end{eqnarray}

\noindent where ${\mathbf{E}}_{ \bot e} = E_{xe}{\mathbf{e}}_{x} + E_{ye}
{\mathbf{e}}_{y} $, ${\mathbf{v}}_{ \bot} = v_{x}{\mathbf{e}}_{x} +
v_{y}{\mathbf{e}}_{y} $, $\beta = {\mathbf{v}}/c$
and $t^\prime $ - is a retarded
time: $c(t -
t') = \left| z - Z(t') \right|$.
The resulting radiation field ${\mathbf{E_{s}}}$, ${\mathbf{H_{s}}}$
for the thin charged
layer is the sum of radiation fields of all electron sheets
(note that the retarded time $t^\prime $ is different for different
sheets). Interaction of each electron sheet with the radiation
field of the layer results in appearance of self-action viscous
force ${\mathbf{F}}_{s}
= e{\mathbf{E}}_{s} + e\left[\beta ,{\mathbf{H}}_{s}
\right]$, which modifies the dynamics of the
sheet. This force is analogous to the Dirac force acting on the
moving electron.
\par
The equations of motion for the electrons in the sheets
have now the following form
\par
\begin{equation}
\label{eq3} {{d\mathbf{p}}\over {dt}} = e{\mathbf{E}} + e\left[
\beta ,{\mathbf{H}} \right] + {\mathbf{F}}_{s} ,
\end{equation}
\noindent where $e$ is the charge of the electron, $\mathbf{p}$ is
relativistic momentum of electrons, $\mathbf{E}$ and $\mathbf{H}$
are the external fields (it is supposed that these fields support
the geometry of electron medium). It is worth to mention that the
radiation reaction force $\mathbf{F_{s}}$ have not only the
transverse component but the longitudinal component as well the
last having essentially nonlinear character. For infinitely thin
electron sheet the self action force is
\begin{equation}
{\mathbf{F}}_{s \bot}  = -2\pi \sigma e
\beta _{ \bot}, \qquad \qquad F_{sz} = - 2\pi \sigma
e\beta _{
\bot} ^{2}\beta _{z}/(1 - \beta _{z}^{2}) ,
\end{equation}

\par
Let now consider the ultraintense plane electromagnetic
wave with frequency $\omega $ falling normally at the layer (wave vector
$k$ is parallel to the $Oz$ axis) so that an acceleration parameter
of the wave $\alpha _{0}=eE_{0}/(m\omega c) >> 1$, where $m$ is  the mass of an
electron, $E_{0}$ is the amplitude of the wave field. Then the
electrons of the layer will accelerate in the $z$ direction to
ultrarelativistic velocities just by the first half wave keeping
initial geometry of the bunch [16].
\par
Account for the layer radiation friction force give some
interesting features to the motion of the electrons inside the
layer. First of all an additional accelerating force emerges which
constitutes the mean value of the Lorentz force. Actually in
this situation the mean Lorentz force is nonzero due to the
extra phase shift between the electromagnetic wave and the
electrons' velocities arising from the action of radiation
friction force (scattering of incident wave). At fig. 1 the
light pressure force acting on the electron layer is presented
for the motion in the given field (fig. 1a, $\alpha =0$) and with
account of the radiation losses (fig. 1b, $\alpha =0.1$), where the
parameter $\alpha  = 2\pi \sigma e/(m\omega c)$ characterizes the
electron density of
the bunch. The amplitude of the external wave is not large
$\alpha _{0}=2$ besides the coulomb interaction of the electrons is
omitted. Fig. 1 demonstrates that the mean Lorents force is not
equal to zero: the deviation of the line to the up is larger
than to the down. Besides the increasing period of the force
corresponds to the increasing longitudinal velocity of the
electron layer so that the effective interaction time of the
layer with each half wave is also increasing. For the motion in
the given field the frequency of Lorentz force and the mean
velocity of electron layer are constants.
\par
Another peculiarity due to account of the radiation friction force
is the emergence of the bunching forces which compress the layer
in the $z$ direction and support its initial geometry [16]. These
forces slow down the electron sheets with velocities larger than
the mean velocity of the layer and accelerate the delaying sheets
(fig. 2). At fig. 2a the bunching effect of the radiation friction
force is not taken into account and the coulomb forces tear the
layer during very short time. At fig. 2b the coulomb forces and
the bunching forces are taken into account simultaneously so the
layer is stable for considerably longer time. The stability of the
layer geometry depends on the value of accelerating parameter
$\alpha _{0}$, initial thickness of the layer and a value of
electron density inside the bunch. Such bunching forces can
compensate partially the coulomb spread of the layer along the $z$
direction (increase of the bunch thickness) and appear to be in a
sense analogous to the magnetic attraction forces between two
parallel currents formed by the moving charges.
\par
One has to account for an action of an ion background of the
target for proper simulation of the process of interaction between
electromagnetic wave and dense plasma layer in the 1D3V model
[16]. Actually this model is adequate in case when the distance
between the electron and ion layers is considerably smaller than
the transverse dimensions of the layers. In our simulations below
the ion background is supposed to be motionless producing only the
Coulomb force acting on the electron layer.
\par
The initial thickness of the target used in simulations is
$l=10^{-2}\mu $ and considerably smaller than the wavelength of the incident
radiation which is $\lambda =1\mu $. The targets with such thickness can be
easily obtained experimentally [20] besides the front of
ultraintense laser pulse squeezes the electron layer in the $z$
direction making it thickness considerably smaller than initial
thickness of the target [16,21].
\par
At fig. 3 the results of computer simulations for the process of
electron layer acceleration are presented (the parameter $\alpha =
1$ in the fig. 3a and $\alpha  = 0.001$ in the fig. 3b). At the
upper plots the dependence of transverse momenta $p_{y}$ for some
electron sheets of the bunch on dimensionless laboratory time
$\omega t$ are presented and at the lower plots the longitudinal
momenta $p_{z}$ are shown. The acceleration parameter $\alpha
_{0}=100$. So all electron sheets of the layer can in principle
move synchronously during acceleration by the first half cycle of
electromagnetic wave constituting the relativistic electron
mirror. For other half cycles of the incident laser pulse the
bunch can be destroyed for large parameter $\alpha $ by the
coulomb forces however for small $\alpha $ values the bunch can be
stable during several half cycles that in laboratory frame can
correspond to the time of several hundreds of femtosecond. So in
the process of interaction of ultraintense femtosecond laser pulse
with thin solid target the relativistic electron mirror can be
formed allowing to realize the up-conversion of light into $X$-ray
and gamma-ray bands.
\par
%\medskip

\section{Up-conversion of counter propagating wave}

\par
%\medskip
Let now consider the simulation of the wave transformation for
weak counter-propagating wave with an amplitude $E_{1}$ ($\alpha
_{1}=eE_{1}/(m\omega c)$) and frequency $\omega_0 $ (for
simplicity we suppose that the frequencies of accelerating and
counter-propagating waves are equal). At fig. 4a the frequency
transformation factor $f_{v}=(1+\beta _{z})/(1-\beta _{z})$ is
presented. The plot of $p_{y}$ in the presence of counter
propagating wave is shown at fig. 4b. The parameter $\alpha
_{1}=1$, the front of this pulse have a delay with respect to
accelerating pulse for the electrons can achieve high energy and
become ultra relativistic. The zoom of electron oscillations at
the top of fig. 4b is presented at fig. 4c. It is worth to mention
that all electron sheets move practically along one trajectory so
the dependencies of $p_{y}$ on time for different sheets are
coincide.
\par
At fig. 5a the reflected radiation of the counter-propagating wave
is presented for $\alpha =0.1$, $\alpha _{0}=100$ and $\alpha
_{1}=1$. Due to the time dependence of transformation factor
$f_{v}$ the frequency of reflected wave have to be altering being
small at the beginning of the acceleration process then largest at
the top of the first half wave and small again at the end of the
first half wave.
\par
At fig. 5b the zoom of the reflected field is shown
(sample 2). The deviation of the radiation field from
sinusoidal form results from high harmonics generating during
the reflection of the counter-propagating wave from relativistic
electron bunch and partially from the simulation numerical
errors.
\par
At fig. 5c the spectra of the samples of equal length at the slope
(1) and at the top (2) of the first half wave of accelerating
pulse are presented (the amplitude and the frequency of the
reflected wave are normalized to the amplitude and the frequency
of the incident accelerating wave). The carrying frequencies in
two cases are different due to the different transformation factor
$f_{v}$ (cf. fig. 4a), besides the spectral bandwidth for sample 1
is larger than for sample 2 because of the faster change of
frequency at the slope.
\par
The amplitude of the reflected wave depends on the values of
parameters $\alpha $ and $\alpha _{0}$. For smaller value of
parameter $\alpha $ the amplitude
of the reflected wave is smaller because of the lower electron
density in the bunch and correspondingly low value of reflection
coefficient. On the other hand the frequency transformation
coefficient is larger for small $\alpha $ value the parameter
$\alpha _{0}$ being
constant because of smaller radiation friction (cf. fig. 3).
\par
For $\alpha _{0}=30$ that can be realized in modern experiments the
maximum frequency transformation factor $f_{v}$ can be about
$4\alpha ^{2}_{0} \approx  2000$
so the reflection of incident wave with wavelength $\lambda =200 nm$ can give
the coherent radiation from the $X$-ray band. For petawatt lasers the
acceleration parameter $\alpha _{0}$ can be about $100\div 200$
so the reflected
coherent radiation can be already in the $\gamma $-ray band.
\par
%\medskip
%\begin{thebibliography}{99}
\section*{References}

\noindent 1. R. C. Elton. $X$-ray lasers, Academic Press, New York, 1990.
\par
\noindent 2. V. I. Vysotski, R. N. Kuzmin. Gamma lasers. Moscow University
Press, Moscow, 1989 (in russian).
\par
\noindent 3. P. Dobiasch, P. Meystre, M. O. Scully. IEEE J. of
Quantum Electron., QE-19, N 12, 1812 (1983).
\par
\noindent 4. A. Loeb, S. Eliezer. Phys. Rev. Lett., 56, 2252 (1986).
\par
\noindent 5. M. Cornacchia. In Proc. SPIE, 3614, Free-Electron Laser
Challenges II, ed. H. E. Bennett, D. H. Dowell, 109 (1999).
\par
\noindent 6. J. Zhang et. al. Science, 276, 1097 (1997).
\par
\noindent 7. B. R. Benware et. al. Phys. Rev. Lett., 81, 5804 (1998).
\par
\noindent 8. A. Goltsov et. al. Plasma Phys. Control Fusion, 41,
A595 (1999).
\par
\noindent 9. Z. Chang et. al. Phys. Rev. Lett., 79, 2967 (1997).
\par
\noindent 10. M. Schnurer et. al. Phys. Rev. Lett., 80, 3236 (1998).
\par
\noindent 11. M. Zepf et. al. Phys. Rev. E, 58, R5253 (1998).
\par
\noindent 12. A. Tarasevitch et. al. Phys. Rev. A, 62, 023816 (2000).

\noindent 13. S. V. Bulanov, F. Califano, G. I. Dudnikova et. al., Problems of
Plasma Theory, ed. V. D. Shafranov, Kluwer Ac. Press (2001).
\par
\noindent 14. G. Mourou, M. D. Perry. Science, 264, 917 (1994).
\par
\noindent 15. K. Landecker. Phys. Rev. 86, 852 (1952).
\par
\noindent 16. V. A. Cherepenin, A. S. Il'in, V. V. Kulagin. Submitted to
Plasma Physics (in russian).
\par
\noindent 17. C. K. Birdsall, A. B. Langdon. Plasma Physics via computer
simulation. Mac Graw-Hill Book Company, 1985.
\par
\noindent 18. A. S. Il'in, V. V. Kulagin, V. A. Cherepenin. Journ. of
Communications Technology and Electronics, 44, 389 (1999).
\par
\noindent 19. V. L. Bratman, S. V. Samsonov. Phys. Lett. A, 206, 377 (1995).
\par
\noindent 20. R. V. Volkov et. al. Quantum Electronics, 24, 1114 (1997).
\par
\noindent 21. B. Rau, T. Tajima, H. Hojo. Phys. Rev. Lett., 78, 3310 (1997).

%\end{thebibliography}

\par
\pagebreak

\medskip
\centerline{Captions for the figures
}
\medskip
\noindent Fig. 1. Force acting on the electron sheet for $\alpha _{0}=2$ in the
given field (a, $\alpha =0$) and in account of radiation losses (b,
$\alpha =0.1$).
\par
\medskip
\noindent Fig. 2. Longitudinal momenta $p_{z}$ for several electron sheets
($\alpha _{0}=100$, $\alpha =1$) with accounting of the coulomb forces only (a)
and the radiation friction forces and the coulomb forces (b).
\par
\medskip
\noindent Fig. 3. Transverse $p_{y}$ and longitudinal $p_{z}$
momenta for several electron sheets for acceleration of the
electron layer in vacuum by the intense electromagnetic wave
($\alpha _{0}=100$): $\alpha =1$ for fig. 3a and $\alpha =0.001$
for fig. 3b. The thickness of the layer in the $z$ direction is
considerably smaller than $\lambda $ for both cases.
\par
\medskip
\noindent Fig. 4. Frequency transformation factor (a), transverse
momentum (b) and the zoom of electron momentum oscillations (c)
for the small counter-propagating wave falling at the electron
mirror. The trajectories of all electron sheets practically
coincide. The counter-propagating wave strikes the electron mirror
at $\omega t=1570$.
\par
\medskip
\noindent Fig. 5. The reflected field (a), zoom of the field (b) and the
spectra (c) of two samples 1 and 2 from different regions of
the accelerating curve.
\par

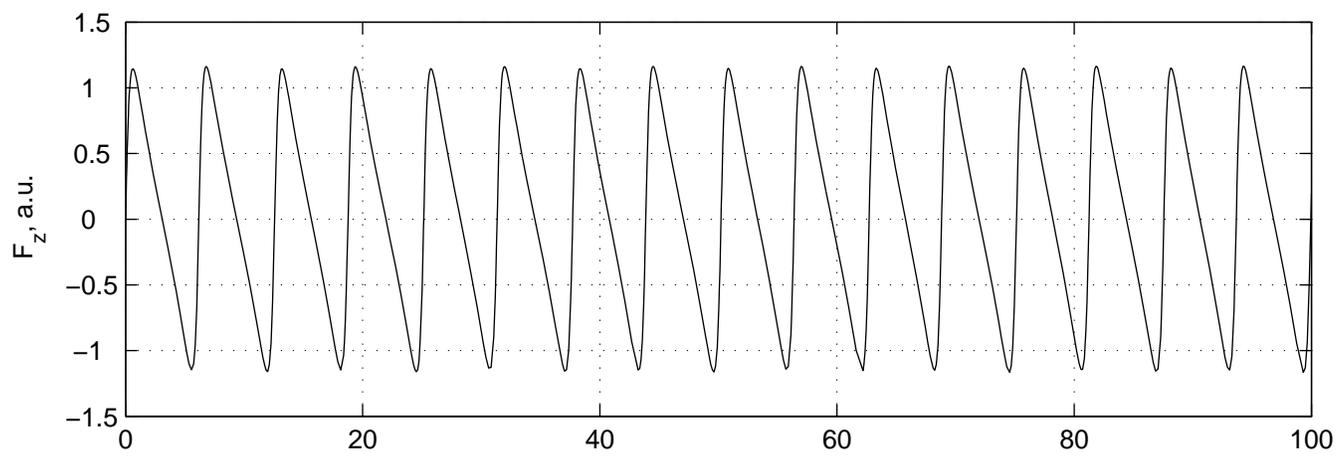
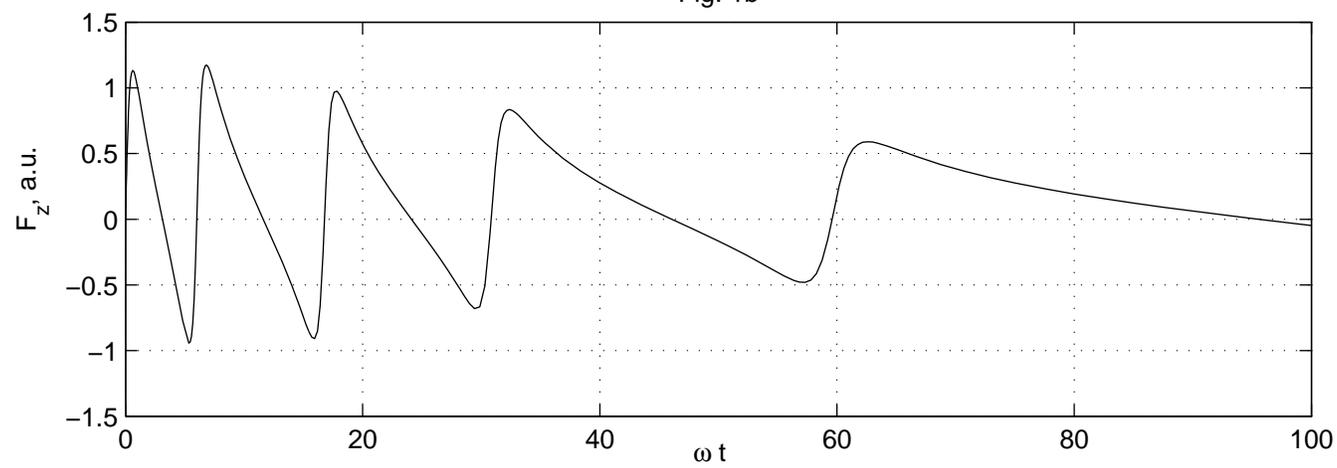
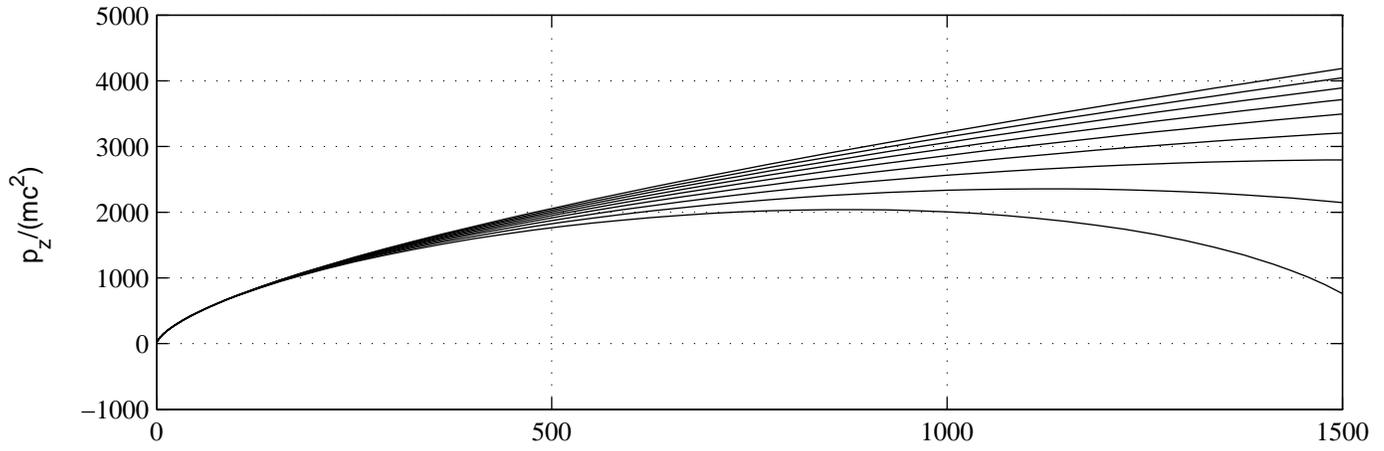
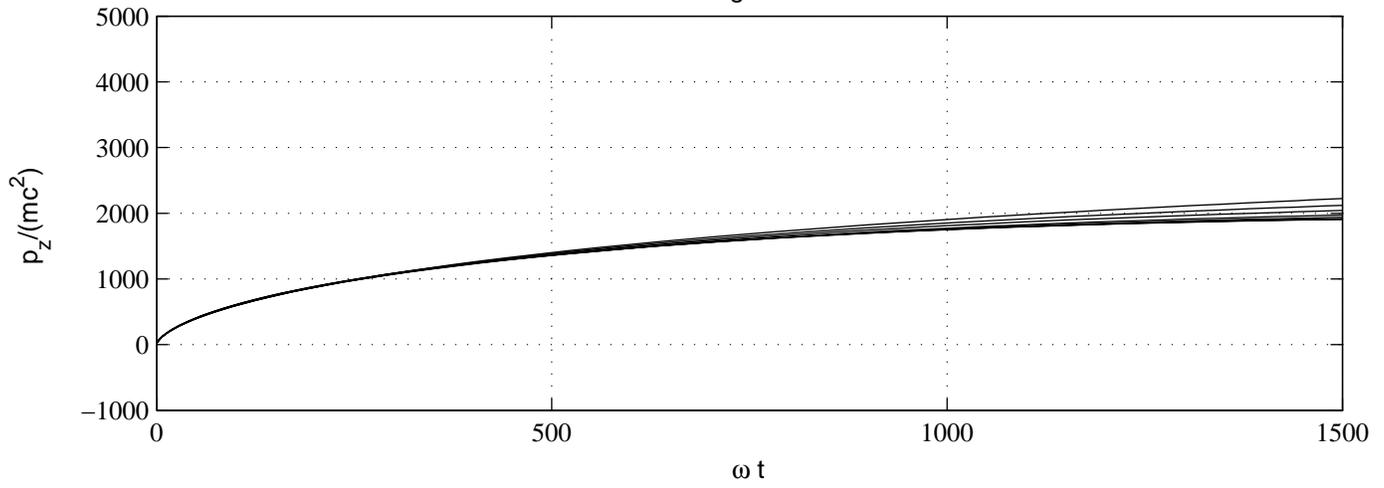
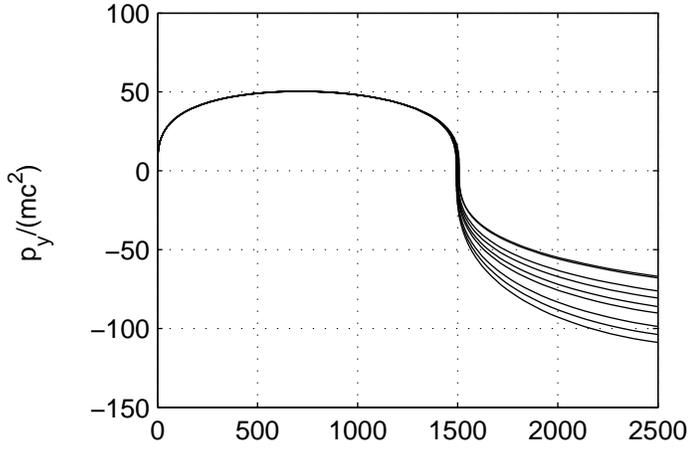
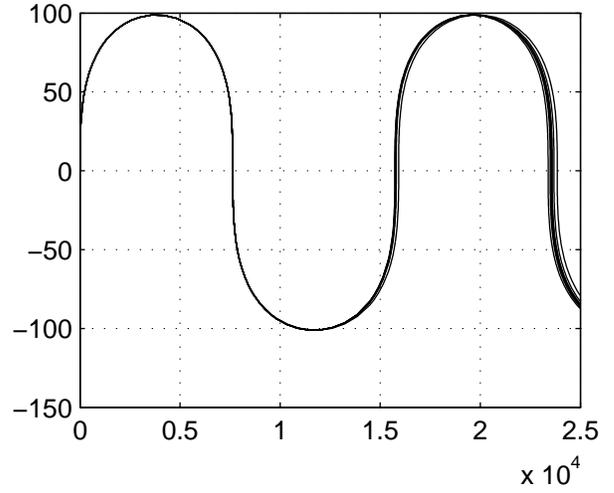
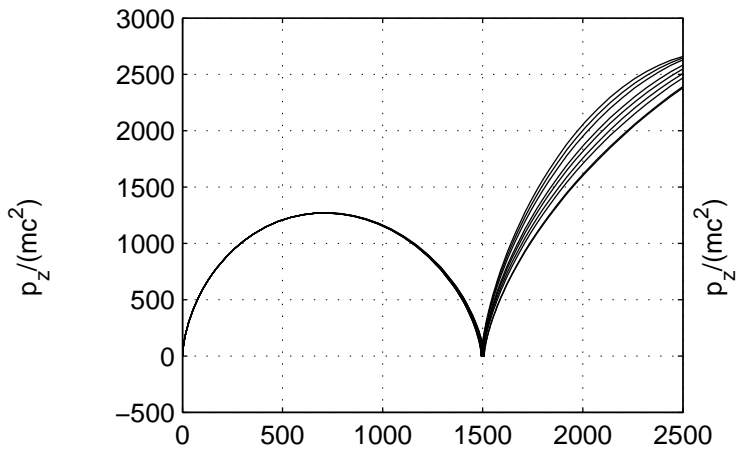
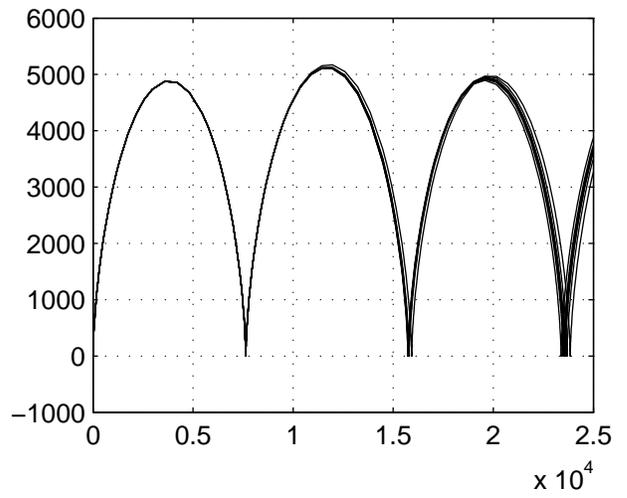
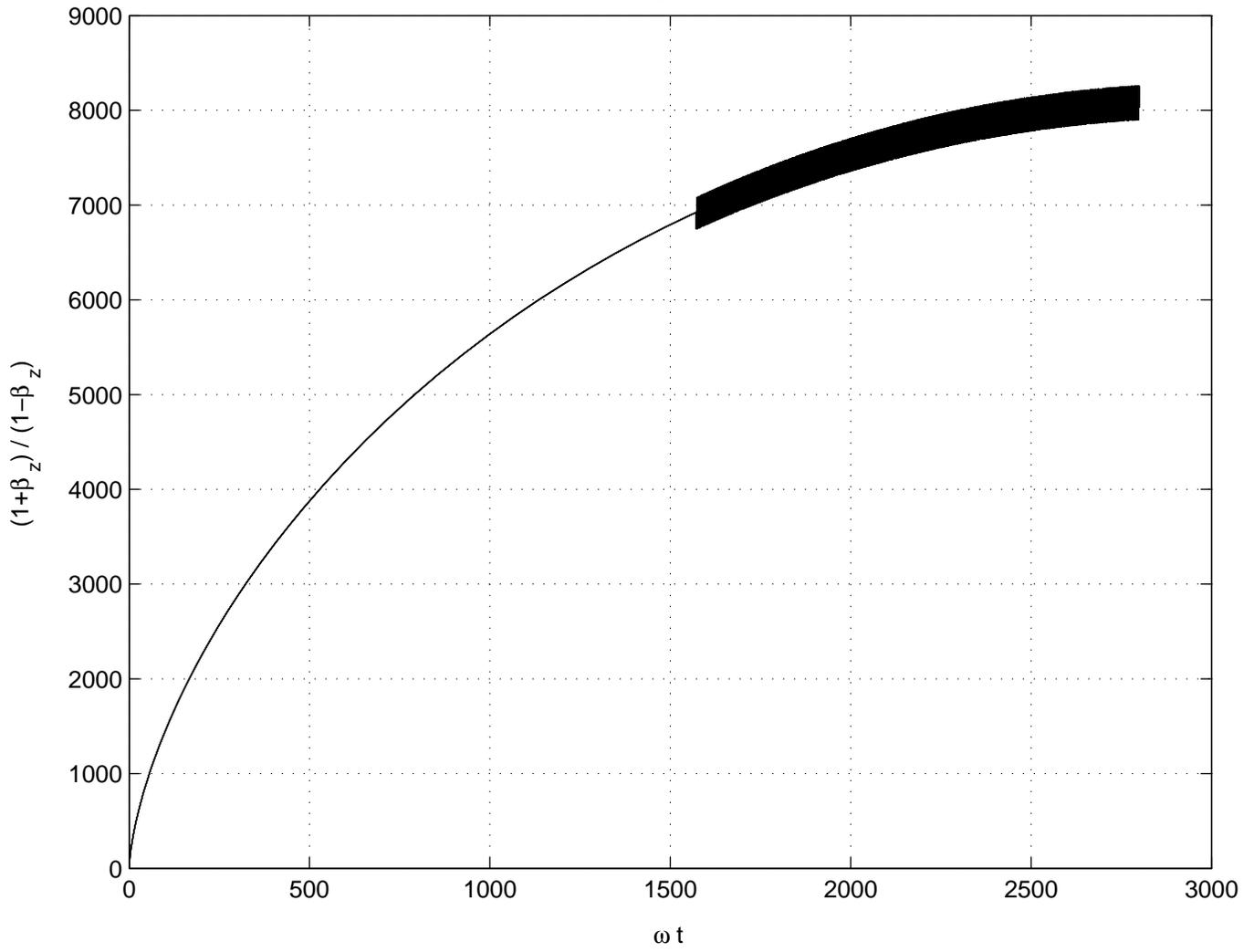
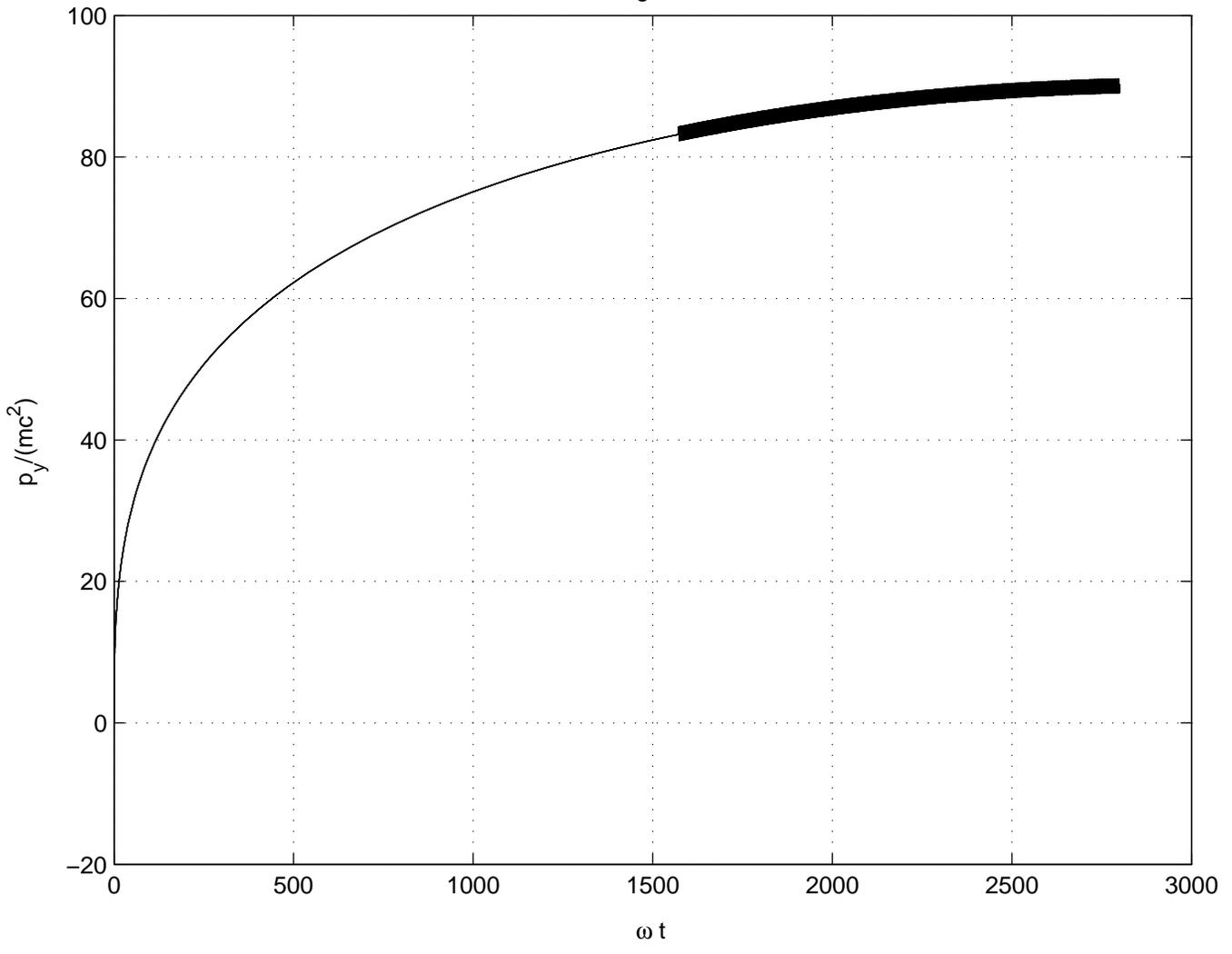
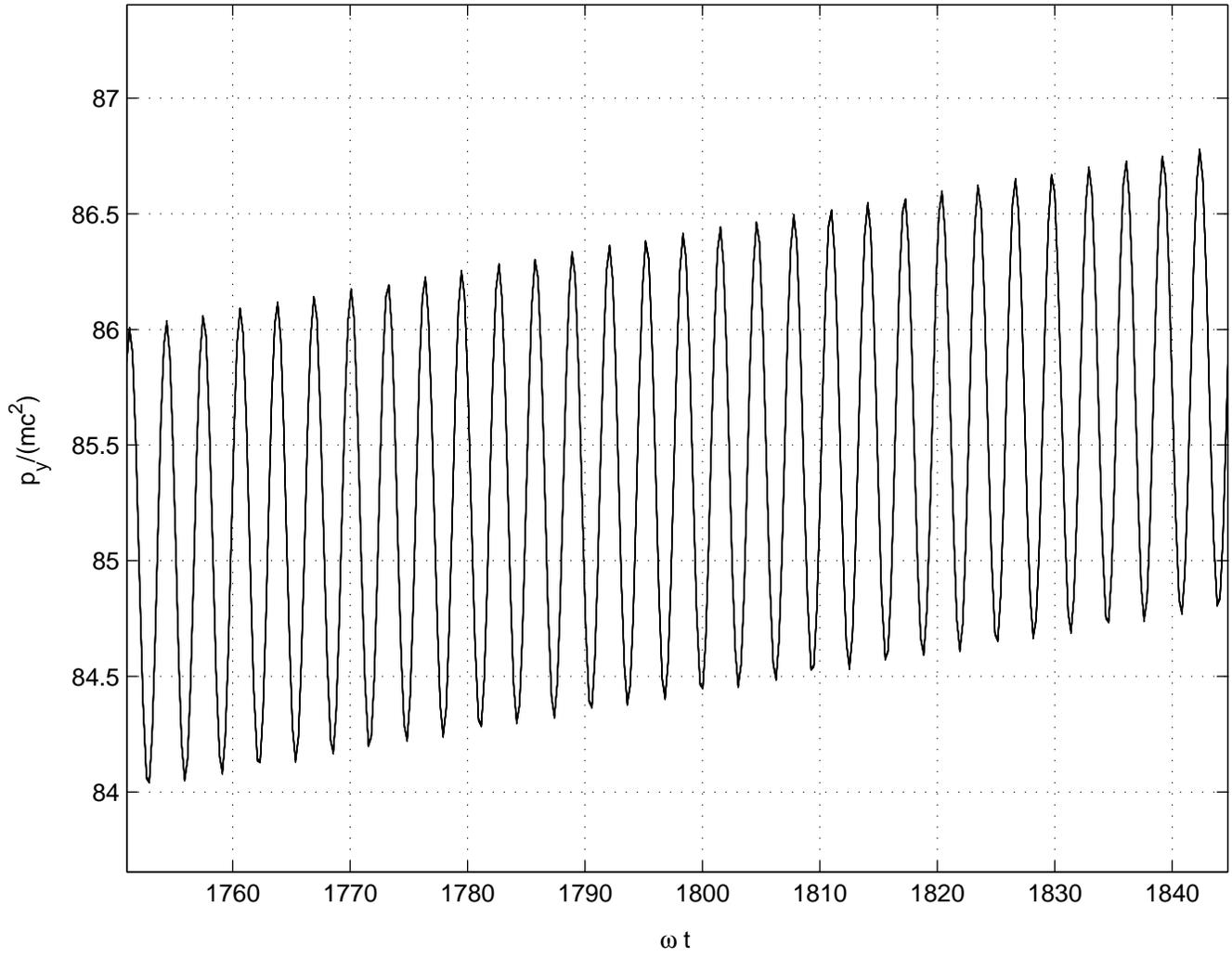
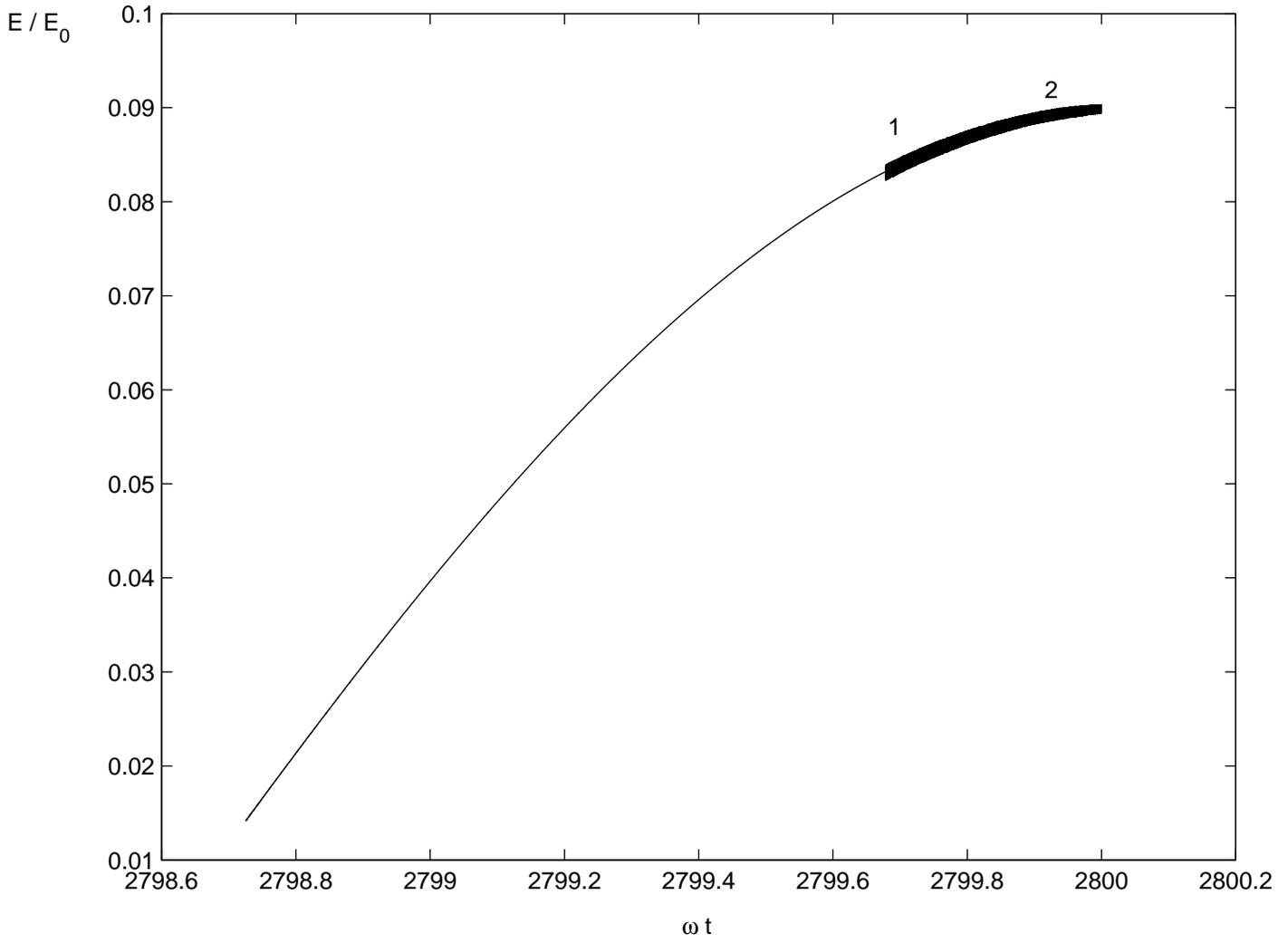
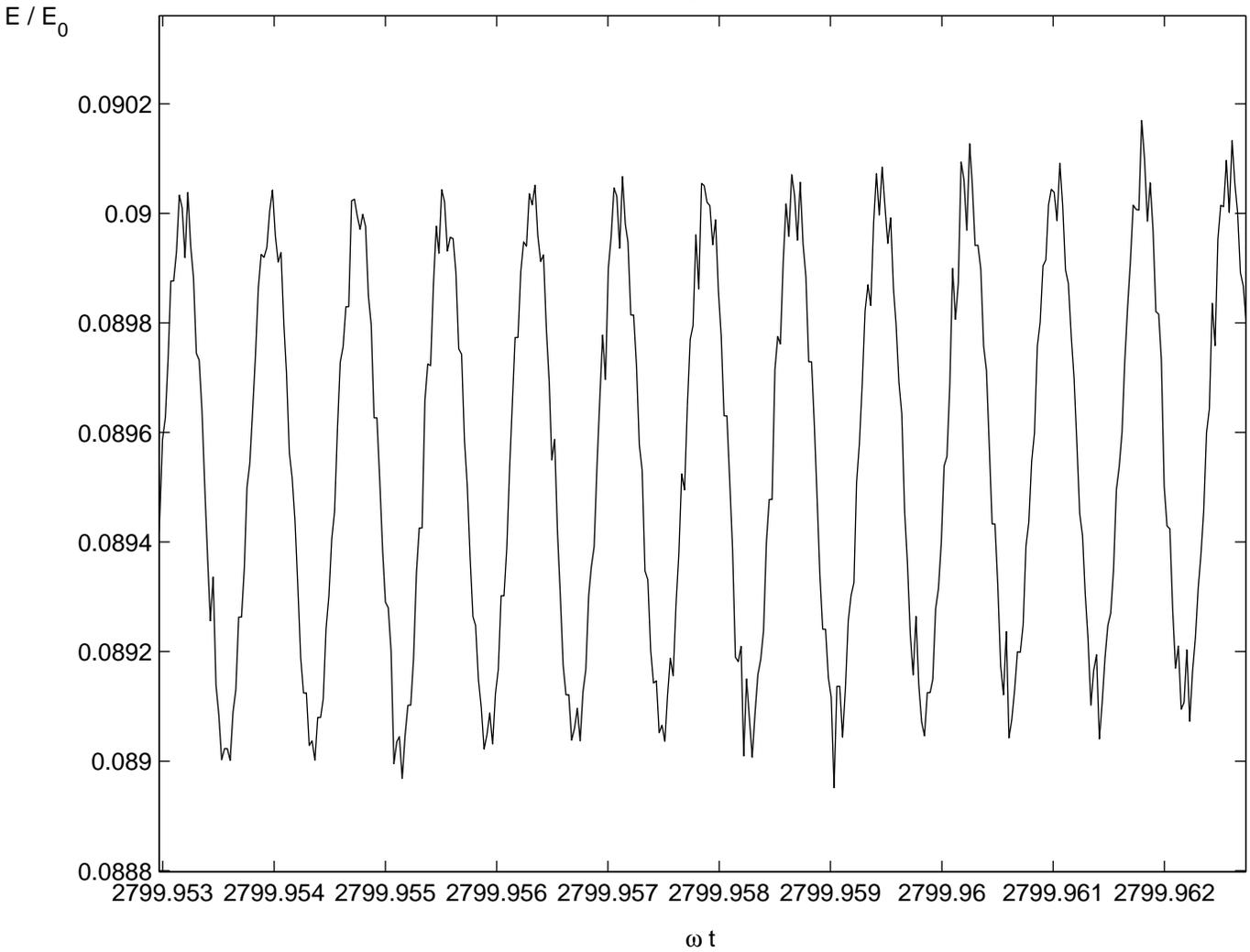
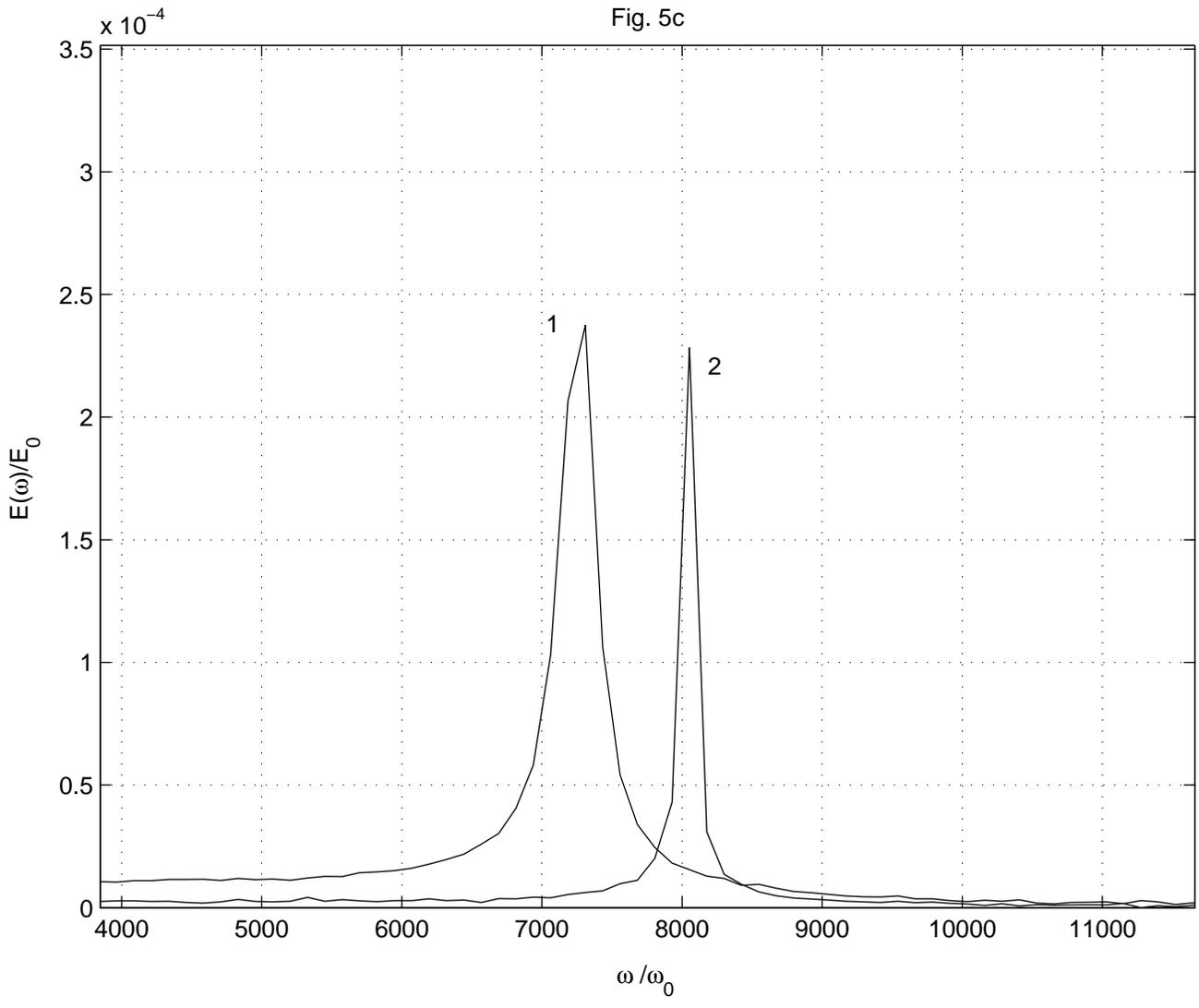
\end{document}